\documentclass
[aps,prb,twocolumn,superscriptaddress,showpacs,floatfix]{revtex4}%
\usepackage{amsfonts}
\usepackage{amssymb}
\usepackage{amsmath}
\usepackage{graphicx}%
\providecommand{\U}[1]{\protect\rule{.1in}{.1in}}
\begin{document}
\title{ Competition between Electron-Phonon coupling and Spin Fluctuations in
superconducting hole-doped BiOCuS}
\author{Luciano Ortenzi}
\affiliation{Max-Planck-Institut f\"{u}r Festk\"{o}rperforschung,
Heisenbergstra$\mathrm{\beta}$e 1, D-70569 Stuttgart, Germany}
\author{Silke Biermann}
\affiliation{Centre de Physique Th{\'e}orique, Ecole Polytechnique, CNRS-UMR7644, 91128
Palaiseau, France}
\author{Ole Krogh Andersen}
\affiliation{Max-Planck-Institut f\"{u}r Festk\"{o}rperforschung,
Heisenbergstra$\mathrm{\beta}$e 1, D-70569 Stuttgart, Germany}
\author{I.I. Mazin}
\affiliation{Naval Research Laboratory, 4555 Overlook Avenue SW, Washington, DC 20375, USA}
\author{Lilia Boeri}
\affiliation{Max-Planck-Institut f\"{u}r Festk\"{o}rperforschung,
Heisenbergstra$\mathrm{\beta}$e 1, D-70569 Stuttgart, Germany}
\date{\today}

\begin{abstract}
BiOCuS is a band insulator that becomes metallic upon hole doping.
Superconductivity was recently reported in doped BiOCu$_{1-x}$S and attributed 
to spin fluctuations as a pairing
mechanism. Based on first principles calculations of the electron-phonon
coupling, we argue that the latter is very strong in this material, 
and probably
drives superconductivity, which is however strongly depressed by the 
proximity to
magnetism. We find however that BiOCu$_{1-x}$S is a quite unique
compound where both a conventional phonon-driven 
and an unconventional triplet superconductivity are possible, and
compete with each other.
We argue that, in this material, it should be possible to
 switch from conventional to unconventional
superconductivity by varying such
parameters as doping or pressure.

\end{abstract}

\maketitle


%
The study of spin fluctuations as superconducting mediators dates back to the
sixties;~\cite{TH:Berk,TH:FayAppel} however, in contrast to the
electron-phonon (EP) interaction, for which a detailed first-principles theory has been
developed in the last twenty years, 
a quantitative theory is still lacking. 
In several materials where at
some point ferromagnetic spin fluctuations (paramagnons) were considered as potential pairing
agents, such as ZrZn$_{2}$,~\cite{zrzn2:DFT} MgCNi$_{3}$,~\cite{mgcni3:DFT} or
Pd metal,~\cite{TH:Berk} phonon and spin fluctuations contributions either cancel, rendering
the material non-superconducting (ZrZn$_{2},$ Pd), or the latter substantially
decreases the superconducting transition temperature. 

Recently, superconductivity with $T_{c}$=5.8 K has been discovered in 
hole-doped BiOCu$_{0.9}$S.~\cite{BSCO:exp:ubaldini}
BiOCuS crystallizes in the {\em ZrCuAsSi}-type structure,
isostructural to the 
1111-family of
Fe-based superconductors,
%
with Cu-S layers playing the role of Fe-As layers.
While Cu-S hybridized $dp$ bands $per$ $se$ are rather similar to
the Fe-As bands in Fe-pnictides, 
the different electronic filling
brings about very different properties in the two systems.
The stoichiometric
 BiOCuS is in fact a band insulator with the Cu being in the $d^{10}$
electronic configuration.~\cite{BSCO:exp:hiramatsu,BSCO:exp:ubaldini, BSCO:exp:anand}
With hole-doping it displays both
a strong tendency to itinerant (ferro)magnetism, and a spectacularly
strong EP coupling, hinting to unconventional, 
triplet $p$-wave,~\cite{BSCO:DFT:mazin}
and conventional, singlet $s$-wave superconductivity, respectively.

In this paper,
we study the interplay between these competing instabilities,
using first-principles calculations of BiOCu$_{1-x}$S as a
function of doping and Stoner parameter, which we use as a proxy for the
tendency to magnetism.
 We find that, as the EP coupling is spectacularly strong, 
it is likely that a conventional superconductivity, even though depressed
by spin fluctuations, is more stable than an unconventional ($e.g.$ $p-$wave)
one.
It appears though that a small variation of parameters can reverse the 
situation and bring triplet superconductivity or long-range magnetism.
We identify two large regions in the parameter space where, respectively, 
ferromagnetism (FM) or conventional $s-$wave superconductivity are the ground
states, with an intermediate region where no FM long range order
is predicted, yet spin fluctuations are strong enough to destroy the s-wave
superconductivity and possibly stabilize a triplet state.

We perform calculations in the linear-response approximation 
for the EP interaction, and in the
local spin density functional version of the random-phase approximation (RPA)
for spin fluctuations, as described below; doping is treated in the rigid
band approximation (RBA).~\cite{technical} 

\begin{figure}[t]
\centerline{\includegraphics*[width=0.95 \columnwidth]{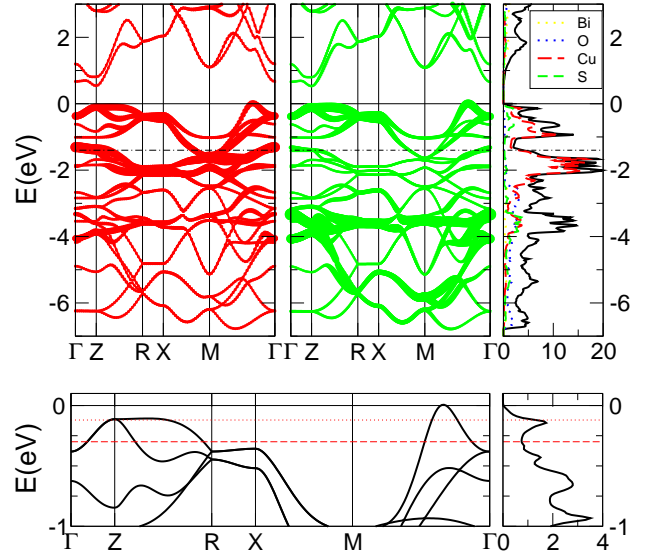}}\caption{(color
  online). \emph{top}
Band structure of BiOCuS, shaded according to the partial Cu $d_{xz+yz}$
(\emph{left}) and S $p_{x+y}$ (\emph{right}) characters: the continuous and 
dashed-dotted lines mark respectively
the position of the Fermi level in the undoped 
compound and that corresponding to the filling  $d^6$
of Fe-pnictides ({\em see text});
the corresponding DOS is also shown. 
\emph{bottom}: a blow-up of the low-energy band structure; 
the dashed and dotted line mark the position of the Fermi
level, corresponding to a hole doping $x=0.1$ and $x=0.5$, 
in RBA.}%
\label{fig:fig1}%
\end{figure}

 The generalized-gradient approximation (GGA) 
band structure and partial electronic density of states (DOS) are shown
%
in Fig.~\ref{fig:fig1}; in agreement with previous calculations,~\cite{BSCO:exp:ubaldini,BSCO:DFT:mazin, BSCO:DFT:shein} we find that the
stoichiometric compound is a semiconductor~\cite{BSCO:exp:ubaldini,
BSCO:exp:hiramatsu} with an indirect gap of
$\Delta\approx$ 0.5 eV (GGA); the top of
the valence band occurs along the $\Gamma-M$ line, and we choose it as the
zero of the energy in the following.
The electronic structure in an energy range $\sim 7$ eV below the top
of the valence band in BiOCuS is derived from Cu $d$ and S $p$
states (see top panel of Fig.~\ref{fig:fig1}).
The Cu $d$ states are centered around $\sim -2$
eV. They hybridize strongly with the S $p$ states, forming
antibonding bands within $\sim 1$ eV below the semiconducting gap. 
The EP matrix element is large for
these bands, as the electronic states 
are very sensitive to ionic displacements.
On the contrary, the deeper, non-bonding, Cu $d$ bands,
centered around $\sim -3$ eV,
are less sensitive to the Cu-S hopping parameters and exhibit
a much weaker EP interaction.
The tendency to magnetism 
is instead strong throughout the entire Cu $d$ band, 
since the Stoner
parameter of Cu is large ($I_{Cu}\approx $0.9 eV).

In pure BiOCuS, Cu is in a nominal $d^{10}$ state and thus not magnetic.
Doping with holes, for $x \le 0.5$, 
shifts the Fermi level into a doubly-degenerate band, with 
dominant Cu $d_{xz}$, $d_{yz}$ and S $p_{x}$, $p_{y}$ characters 
and large DOS.
The bottom panel of Fig.~\ref{fig:fig1} shows a blow-up of the band
structure in the energy range relevant for superconductivity.
The dotted and dashed lines
indicate respectively 
the position of the Fermi level for $x=0.1$, corresponding 
to the doping for which superconductivity was observed
in Ref.~\onlinecite{BSCO:exp:ubaldini}, and
$x=0.5$, which is the highest doping considered in our RBA study.

If we could shift the Fermi level further down,
so as to cut the band structure at
$\sim -1.4$ eV (dash-dotted line in the top panel of Fig.~\ref{fig:fig1}),
we would find a striking similarity with the familiar 
low-energy electronic structure of Fe-pnictides, with the $xz,yz$ hole
and electron pockets, centered at $\Gamma$ and $M$ respectively.
The
DOS and the $p-d$ hybridization here are small, thus the tendency to antiferro-
(rather than ferro-) magnetism, and low EP interaction.
This is indeed what first-principles calculations
find in Fe-pnictides.~\cite{FESC:boeri:eph}

We now go back to discuss the behavior of BiOCu$_{1-x}$S for 
$x \le 0.5$, using the bottom panel of Fig.~\ref{fig:fig1}.
For $x \ge 0.1$, 
we find 
the ground state of the system is FM,
 both in the local spin density approximation (LSDA) 
and in the GGA.~\cite{BSCO:DFT:mazin}
This can be understood in terms of the Stoner criterion for FM, $IN_0>1$.
When holes are introduced into the system, and the Cu charge state is being
reduced from $d^{10}$ to $d^{9}$,
the Fermi level moves into a flat region
of the band structure, which gives rise to a high peak in the DOS
($N_0=2.1$ st/eV spin).
 Since the Stoner parameter of atomic Cu is $I_{Cu}\approx0.9$ eV, the
latter value of the DOS is well above the Stoner criterion for FM, $IN_0>1$.
In BiOCu$_{1-x}$S the actual value of $I \le I_{Cu}$, due to the Cu-S 
hybridization. It can be estimated from the splitting
 $\Delta E=m I$ between majority and minority bands in the FM state,
 where $m$ is the value of the self-consistent magnetic moment.
We find $m\lesssim 0.1$ for all dopings considered, and $I=0.53$ eV in LSDA 
and $I=0.67$ eV in GGA, independent of doping.
%
%
So far, however, experiments have seen no trace of static 
magnetism; this is consistent with the tendency of LSDA calculations to 
overestimate the tendency to itinerant magnetism
with respect to experiment
in the vicinity of a magnetic quantum critical point (QCP), where the system
exhibits strong spin fluctuations.~\cite{mazin:QCP} We will return to this
issue in more detail, after discussing the results for the EP interaction.

\begin{figure}[t]
\centerline{\includegraphics*[width=0.8 \columnwidth]{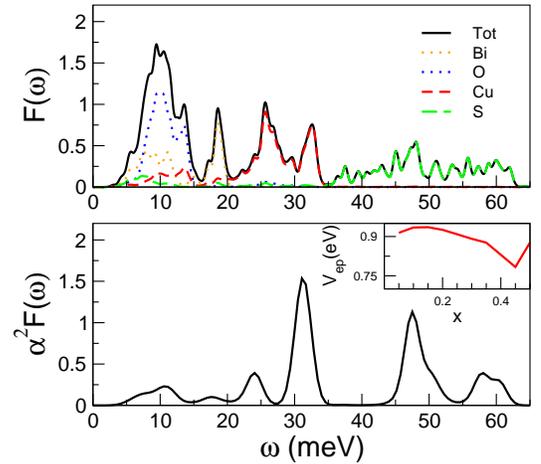}}\caption{(color
  online).
From top to bottom: Partial Phonon density of States (PDOS), Eliashberg
spectral function for $x=0.1$, in RBA, and (inset)
ratio between the coupling constant and the DOS as a function of doping. 
}%
\label{fig:fig2}%
\end{figure}
Fig.~\ref{fig:fig2} summarizes the EP properties of the hole-doped BiOCu$_{1-x}$S. The 
partial phonon density of states (PDOS) of the undoped
compound extends up to 65 meV; vibrations of the
Bi-O layers are concentrated at energies $\leq 20$ meV, while modes involving
the Cu-S layers are found at higher energies. The S atoms give rise to a
very broad feature in the PDOS, from 40 to 65 meV.
Using this phonon spectrum,
we calculate the Eliashberg spectral function $\alpha^{2}F(\omega)$ of the
hole-doped BiOCu$_{1-x}$S:
%
$$
\alpha^{2}F(\omega)=\frac{1}{N_{0}}\sum_{\mathbf{k},\mathbf{q},\nu,n,m}%
\delta(\epsilon_{\mathbf{k}}^{n})\delta(\epsilon_{\mathbf{k+q}}^{m}%
)|g_{\mathbf{k},\mathbf{k+q}}^{\nu,n,m}|^{2}\delta(\omega-\omega
_{\nu\mathbf{q}}),%
$$
evaluating the average of the EP matrix elements $g_{\mathbf{k},\mathbf{k+q}
}^{\nu,n,m}$ on the Fermi surface $\delta(\epsilon_{\mathbf{k}}^{n})$,
obtained by a rigid-band shift corresponding to the doping level. From the
Eliashberg function we calculate the EP coupling constant:\\
$\lambda_{\text{ep}}=2\int_{0}^{\infty}d\Omega\alpha^{2}F(\Omega)/\Omega$.

For all dopings $x \leq 0.5$, we find that only two groups of phonon modes,
corresponding to the out-of-plane vibrations of the Cu-S layers, have sizable
EP matrix elements $g_{\mathbf{k},\mathbf{k+q}}^{\nu,n,m}$: these give rise to
two narrow peaks in $\alpha^{2}F(\omega)$, centered at $33$ meV and $50$ meV.
The lower panel of Fig.~\ref{fig:fig2} shows an example of $\alpha^{2}
F(\omega)$ for $x=0.1$.

Since the shape of the Eliashberg function does not depend on $x$ for all dopings considered, the
total EP coupling depends on doping only through the value of the density of
states at the Fermi level, $N_0$. We thus rewrite $\lambda_{\text{ep}}$ as
%
$\lambda_{\text{ep}}=N_0V_{\text{ep}}$.
As the inset of Fig.~\ref{fig:fig2} shows, $V_{\text{ep}}\simeq0.9$ eV spin
f.u\emph{.}
at all dopings for $x\leq0.5$.
For comparison, $V_{\text{ep}}=0.1$ eV spin f.u. in LaOFeAs and
$V_{\text{ep}}=0.3$ eV spin f.u. in Pd (\emph{i.e.} in metals
where the lattice plays a minor role compared to spin fluctuations) while it
is much larger in good EP superconductors: $V_{\text{ep}}=2.5$ eV spin f.u. in
MgB$_{2}$ or $V_{\text{ep}}=6.6$ eV spin f.u. in Pb.

For $x=0.1$, $N_0=1.93$ st/eV 
spin f.u., $\lambda_{\text{ep}}=1.74$ and the
logarithmically averaged phonon frequency $\omega
_{\text{log}}=263$ K. This EP interaction would then
give rise to a $T_{c}$ of $33$ $K$, assuming a typical value for the Coulomb pseudopotential, $\mu^{\ast}=0.1$.

%
This is much larger than the experimental value 
$T_{c}=5.8$ $K$,~\cite{BSCO:exp:ubaldini}
which would correspond to $\lambda_{\text{ep}}=0.6$.~\cite{foot:Tc}
A factor three discrepancy 
is well above the typical uncertainty of $T_{c}$ in
similar calculations, stemming from the computational uncertainty on
$\lambda_{\text{ep}}$, typically 10\%, or from the uncertainty of $\mu^{\ast}$
. 

The most straightforward explanation, in the present case, is a
suppression of phonon-mediated pairing by strong paramagnons, due to proximity to a FM QCP.
We now estimate this effect, using the RPA.
 Let
$\lambda_{\text{sf}}^{\text{s}}$ be the coupling to spin fluctuations in
the singlet channel; the effect of paramagnons is to suppress 
superconductivity in the singlet channel by
depressing the effective coupling constant ($\lambda_{\Delta}=\lambda
_{\text{ep}}-\lambda_{\text{sf}}^{\text{s}}$) and increasing the effective
mass of the carriers by the factor $1+\lambda_{Z}=1+\lambda_{\text{ep}
}+\lambda_{\text{sf}}^{\text{s}}$. This effect has
been studied in Ref.~\onlinecite{PM:Eliashberg:dolgov} where the following 
expression for $T_{c}$ was derived (and verified by comparison with numerical 
solutions of the Eliashberg equations):
\begin{equation}
T_{c} =\frac{\omega _{\text{log}}}{1.45}\exp
\left\{ \frac{-(1+\lambda_{Z})}{\lambda _{\Delta}-\mu
^{\ast }(1+0.5 \frac{\lambda _{\Delta}}{1+\lambda_{Z}})}
\right\}.
\label{eq:Tc2}
\end{equation}
Here we assume for simplicity that the characteristic 
frequencies of phonons and electrons are the same. 
Eq.~\ref{eq:Tc2} can also be generalized to 
triplet superconductivity, with
the substitution:
$\lambda_{\Delta}\rightarrow\lambda_{\text{sf}}^{\text{t}}$;
$\lambda_{Z}\rightarrow\lambda^{\text{t}}_{Z}=\lambda_{\text{ep}}+\lambda_{\text{sf}}^{\text{t}}$,
where $\lambda_{\text{sf}}^{\text{t}}=\frac{1}{3}\lambda_{\text{sf}}^{\text{s}}$
is the coupling to spin fluctuations in
the triplet channel.~\cite{Scalapino}
%
%
Eq.~\ref{eq:Tc2} gives
an appreciable $T_{c}$ only if 
the denominator in the exponential is positive.
For small $\mu^{\ast}$, this is the case, when $\lambda_{\Delta} > 0$.
We therefore use $\lambda_{\Delta}$
to define the phase diagram of hole-doped BiOCu$_{1-x}$S:
using the RBA, we take $\lambda_{\text{ep}}(x)=V_{\text{ep}}N_0(x)$, where
$N_0(x)$ is the DOS at the Fermi level at doping $x$. For the coupling to spin
fluctuations we use the following expression:
\begin{equation}
\lambda_{\text{sf}}^{\text{s}}(x)=\frac{3}{2}\frac{N_0^{2}(x)I^{2}}%
{1-IN_0(x)}\label{eq:singlet3}%
\end{equation}
where $I$ is, in the LDA parlance, the
Stoner parameter.~\cite{footnote} Eq.~\ref{eq:singlet3} 
is similar to the well-known expression for the 
spin fluctuations induced interaction in the singlet 
channel,~\cite{Scalapino} averaged over the Fermi surface. Note that in the triplet channel
the spin fluctuations interaction is three times smaller, and also the averaging for both
$\lambda_{\text{sf}}$ and $\lambda_{\text{ep}}$
is performed with a weighting factor $\hat{v}_F({\bf k})\cdot \hat{v}_F({\bf k'})$.

\begin{figure}[h]
\centerline{\includegraphics*[width=1.0 \columnwidth]{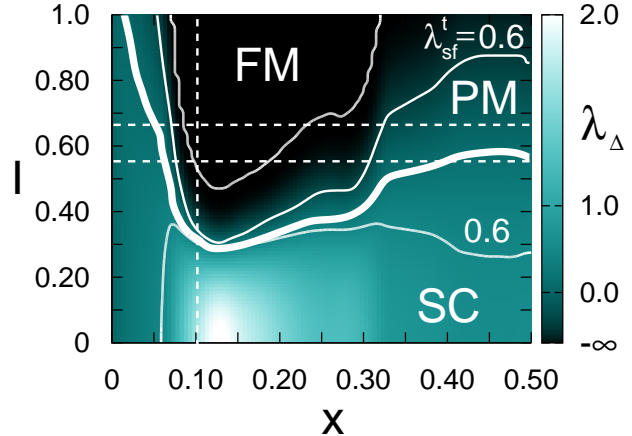}}
\caption{(color online).
Phase diagram of BiOCu$_{1-x}$S, defined by $\lambda_{\Delta}$
 as a function of doping ($x$) and Stoner parameter $I$.
whose value is represented by the color scale.
The two horizontal dashed lines correspond to
$I_{\text{LDA}}$=0.53 eV and $I_{\text{GGA}}$=0.67 eV.
The vertical dashed line indicates the doping for
which superconductivity was observed in
Ref.~\onlinecite{BSCO:exp:ubaldini}.
In the region (FM) the system shows a FM instability, defined by
the condition ($N_0I \ge 1$); elsewhere the system is paramagnetic (PM). 
Below the bold line (which marks the condition
$T^{\text{s}}_c=T^{\text{t}}_c$)
the ground state is a conventional singlet superconductor. Above the bold
line a triplet superconducting state is more stable.
The isolines $\lambda_{\Delta}=0.6$ and $\lambda^{\text{t}}_{\text{sf}}=0.6$
indicate the values of $I,x$ which reproduce the experimental $T_{c}=5.8$ K of
Ref.~\onlinecite{BSCO:exp:ubaldini} in the singlet and triplet channel 
respectively.}
\label{fig:fig3}
\end{figure}

A well-known LDA problem is that, due to its mean-field character,
it overestimates the tendency to static magnetism.~\cite{larson} This can
be corrected by introducing a phenomenological Stoner $I$, reduced from its 
LDA value. In this spirit, in the following, we treat $I$
as a free parameter, and plot the phase diagram of BiOCuS in the $(x,I)$ space.
The results are shown in Fig.~\ref{fig:fig3}.

If $\lambda_{\Delta}\gg\mu^{\ast}$ a conventional EP superconductivity, 
albeit depressed by spin fluctuations, is a stable
zero-temperature ground state. 
As the Stoner parameter is increased $\lambda_{\Delta}$
goes down, and a competing instability against a triplet state emerges
when the critical temperature in the singlet channel 
$T^{\text{s}}_c$, defined by Eq.~\ref{eq:Tc2},
becomes equal to that in the triplet channel ($T^{\text{t}}_c$).
Finally, as the tendency to magnetism is increased even further, the Stoner 
criterion $N_0I>1$ is satisfied, and the system becomes ferromagnetic (Fig. \ref{fig:fig3}).

One can see that, had we used the LDA or GGA value for $I$,
 for dopings close to $x=0.1$, we would have found BiOCuS inside the FM region. 
However, at
$x=0.1$ experiments see no trace of static FM order, a sign of inadequacy of
the mean field character of magnetism in LSDA. Reducing the LDA value of $I$
to $I_{\text{eff}}=0.51$ eV suppresses the magnetic instability at $x=0.1$; a 
reduction to $I_{\text{eff}}=0.39$ eV brings the 
estimated triplet $T_c$ into agreement with the experimental one, and a 
reduction to 0.25 eV does the same with the conventional singlet $T_c$.
For typical itinerant magnets renormalizing $I_{LDA}$ by $\sim 30-40$\% provides 
reasonable agreement with the experimental magnetic susceptibilities,~\cite{larson} in the same ballpark as the reduction introduced above.

In other words, BiOCu$_{0.9}$S is a unique example where a spin fluctuations driven triplet
superconductivity
is nearly degenerate with the phonon-driven singlet superconductivity, and the critical temperature 
is sizable for both symmetries. 
Given that the actual BiOCu$_{0.9}$S samples are rather
dirty, one may conjecture that samples studied in Ref.~\onlinecite{BSCO:exp:ubaldini}
are on the conventional side of the phase diagram, but the fact that 
superconductivity
appears to be so difficult to reproduce may be due to the fact that slightly different samples 
may appear outside of the stability range of singlet pairing in the phase
diagram in Fig.~\ref{fig:fig3}. In principle, one can use pressure and doping, which control 
$I$ and $N_0$ respectively, to move around intentionally in
the proposed phase diagram.

This tunability comes about because of the combination
of two factors: an exceptionally strong EP interaction in the singlet channel that is
essentially canceled out in the triplet channel, and a strong spin fluctuations coupling that
competes with EP interaction in the singlet channel.
The occurrence of these two large coupling constants 
can be seen as the result of three concurring elements:
a strong $d-p$ hybridization, 
that causes large EP matrix elements; 
the large value of the Stoner parameter of Cu,
that causes a strong tendency to magnetism; and,
finally, the presence of a large peak in the electronic DOS,
which favors FM and enhances the coupling
constants for superconductivity both in the singlet and triplet channel.

\textbf{Acknowledgements:} The authors would like to thank 
D. J. Scalapino, D. van der Marel and E. Giannini for
useful discussions, and M. Calandra for help in developing the 
rigid-band routine. O.K.A. and S.B. acknowledge also the hospitality of 
KITP Santa Barbara, where this work was started. 
This research was supported in part by the
National Science Foundation under Grant No. PHY05-51164, the French ANR under
project Correlmat and IDRIS/GENCI under project 101393.



\begin{thebibliography}{99}                                                                                               %
\bibitem {TH:Berk}N. F. Berk and J. R. Schrieffer, Phys. Rev. Lett.
\textbf{17}, 433 (1966).


\bibitem {TH:FayAppel}D. Fay, J. Appel, Phys. Rev. B \textbf{22}, 3173 (1980).



\bibitem {zrzn2:DFT}G. Santi, S. B. Dugdale, and T. Jarlborg, Phys. Rev. Lett.
\textbf{87}, 247004 (2001);
I. I. Mazin and D. J. Singh, Phys. Rev. B 69, 020402(R) (2004).



\bibitem {mgcni3:DFT}
H. Rosner, {\it et al},
Phys. Rev. Lett. \textbf{88}, 027001 (2001);
%
D. J. Singh and I. I. Mazin, Phys. Rev. B \textbf{64}, 140507(R) (2001);
%
A. Yu. Ignatov, S. Y. Savrasov, and T. A. Tyson, Phys. Rev. B \textbf{68},
220504(R) (2003).


\bibitem {BSCO:exp:ubaldini}A. Ubaldini, E. Giannini, C. Senatore, D. van der
Marel, Physica C \textbf{470}, S356-S357 (2010).


\bibitem {BSCO:exp:hiramatsu}H. Hiramatsu,{\it et al},
Chem. Mater. \textbf{20}, 326 (2008).

\bibitem {BSCO:exp:anand}A. Pal, H. Kishan and V.P.S. Awana, J. Supercond.
Novel Magn. \textbf{23}, 301 (2010).


\bibitem {BSCO:DFT:mazin}I.I. Mazin, Phys. Rev. B \textbf{81}, 
140508(R) (2010).



\bibitem {technical}
%
We use the experimental crystal structure, $a$= 3.8726 \AA \ and
$c$=8.5878 \AA , $z_{\text{Bi}}$=0.14829, $z_{\text{S}}$%
=0.671.~\cite{BSCO:exp:ubaldini, BSCO:exp:hiramatsu}
For the band structure and DOS calculations,
we employ the linearly augmented plane wave methods, as
implemented in the Wien2K code.~\cite{wien2k}
The linear response EP calculations are performed in the generalized gradient
approximation~\cite{DFT:PBE} using plane-waves~\cite{DFT:PWSCF},
ultra-soft~\cite{Vanderbilt} and norm-conserving
Martin-Trouillers~\cite{DFT:mt} pseudopotentials.
We employ a cut-off of 100 (800) Ryd for the wave-functions (charge
densities) and $4\times4\times2$ $\mathbf{k}$-mesh for the
the self-consistent cycles, Finer grids ($48\times48\times24$) are used for
evaluating the EP linewidths, and the densities of states (DOS) in the doped
regime. Dynamical matrices and EP linewidths are calculated on a
$8\times8\times2$ uniform grid in $\mathbf{q}$-space. Phonon frequencies
throughout the Brillouin Zone are obtained by Fourier interpolation. The
(perturbed) potentials and charge densities, as well as the phonon
frequencies, are calculated self-consistently at zero doping ($x=0$);
the effect of doping on the EP coupling was then estimated using RBA.

\bibitem {wien2k}http://www.wien2k.at.




\bibitem {DFT:PBE}J. P. Perdew, K. Burke, and M. Ernzerhof, Phys. Rev. Lett.
\textbf{78}, 1396 (1997).

\bibitem {DFT:PWSCF}P. Giannozzi et al., http://www.quantum-espresso.org.

\bibitem {Vanderbilt}D. Vanderbilt, Phys. Rev. B \textbf{41}, R7892 (1990).

\bibitem {DFT:mt}N. Trouiller and J.L. Martins, Phys. Rev. B \textbf{43},
1991, (1993).


\bibitem {BSCO:DFT:shein}I. R. Shein and A. L. Ivanovskii, Solid State Commun.
\textbf{ 150}, 640 (2010).

\bibitem {FESC:boeri:eph}L. Boeri, \emph{et al.},
Phys. Rev. B {\bf 82}, 020506 (2010).

\bibitem {mazin:QCP} ``Density Functional Calculations near Ferromagnetic
Quantum Critical Points", I. I. Mazin, D.J. Singh, and A. Aguayo, in
Proceedings of the NATO ARW on Physics of Spin in Solids: Materials, Methods
and Applications, ed. S. Halilov, Kluwer, 2004.

\bibitem{foot:Tc}
The 
values of $T_c$ are obtained using Eq.~\ref{eq:Tc2},
with $\lambda^{\text{s}}_{\text{sf}}=0$. Using the Mc-Millan formula gives 
differences of less than 1K in $T_c$.

\bibitem {PM:Eliashberg:dolgov}
O. V. Dolgov, {\it et al},
Phys. Rev. Lett. \textbf{95}, 257003 (2005).



\bibitem {Scalapino}D. J. Scalapino J. Low Temp. Phys. \textbf{117}, 179 (1999)




\bibitem{footnote}In BiOCuS, the
LDA spin susceptibility has a large peak at $\mathbf{q}=0$, due to intraband
processes, and four smaller peaks at $\mathbf{\bar{q}}\sim(\mathbf{\pi
}/8,\mathbf{\pi}/8)$, due to interband transitions; the relative weight is
such that $\chi_{0}(\mathbf{0},0)\approx2\chi_{0}(\mathbf{\bar{q}},0)$. Near
the instability it is reasonable to keep only the contribution at
$\mathbf{q}=0$; in the $\omega=0$ limit we obtain Eq.
(\ref{eq:singlet3}).




\bibitem {larson}P. Larson, I. I. Mazin, and D. J. Singh, Phys Rev. B \textbf{69},
064429 (2004); arXiv:cond-mat/0401563.






























































\end{thebibliography}
\end{document}